\documentclass[aps,pre,twocolumn,showpacs,superscriptaddress]{revtex4}

\usepackage{graphicx}
\usepackage{dcolumn}
\usepackage{bm}

\def\be{\begin{equation}}
\def\ee{\end{equation}}
\def\bea{\begin{eqnarray}}
\def\eea{\end{eqnarray}}
\def\bi{\begin{itemize}}
\def\ei{\end{itemize}}

\begin{document}

\title{Universal Non-Gaussian Velocity Distribution in Violent Gravitational
Processes}

\author{Osamu Iguchi \footnote{osamu@phys.ocha.ac.jp}}
  \affiliation{Department of Physics, Ochanomizu University,
	       2-1-1 Ohtuka, Bunkyo, Tokyo 112-8610, Japan}

\author{Yasuhide Sota \footnote{sota@cosmos.phys.ocha.ac.jp}}
  \affiliation{Department of Physics, Ochanomizu University,
	       2-1-1 Ohtuka, Bunkyo, Tokyo 112-8610, Japan}
  \affiliation{Advanced Research Institute for Science and Engineering, 
               Waseda University, Ohkubo, Shinjuku--ku, Tokyo 169-8555, Japan}

\author{Takayuki Tatekawa \footnote{tatekawa@gravity.phys.waseda.ac.jp}}
  \affiliation{Department of Physics, Waseda University,
               3-4-1 Okubo, Shinjuku-ku, Tokyo 169-8555, Japan}

\author{Akika Nakamichi \footnote{akika@astron.pref.gunma.jp}}
  \affiliation{Gunma Astronomical Observatory, 
               6860-86, Nakayama, Takayama, Agatsuma, Gunma 377-0702, Japan}

\author{Masahiro Morikawa \footnote{hiro@phys.ocha.ac.jp}}
  \affiliation{Department of Physics, Ochanomizu University,
               2-1-1 Ohtuka, Bunkyo, Tokyo 112-8610, Japan}

\begin{abstract}
We study the velocity distribution in spherical collapses and 
cluster-pair collisions by use of N-body simulations. 
Reflecting the violent gravitational processes, 
the velocity distribution of the resultant quasi-stationary state 
generally becomes non-Gaussian. 
Through the strong mixing of the violent process, 
there appears a universal non-Gaussian velocity distribution, 
which is a democratic (equal-weighted) superposition of 
many Gaussian distributions (DT distribution). 
This is deeply related with the local virial equilibrium and 
the linear mass-temperature relation which characterize the system. 
We show the robustness of this distribution function 
against various initial conditions which leads to the violent gravitational process.
The DT distribution has a positive correlation with the energy fluctuation of the system.
On the other hand, 
the coherent motion such as the radial motion in the spherical collapse 
and the rotation with the angular momentum suppress 
the appearance of the DT distribution. 
\end{abstract}


\maketitle



\section{Introduction}


Galaxies and clusters of galaxies are typical structures formed through the
gravity of their own. We would like to understand the history and universal
characterization of these self-gravitating structures. Especially, we are
interested in how extensively the formation process of them are involved in the
resultant universal structure through their gravitational interaction. In
these structure formation of self-gravitating systems (SGS), \textit{a cold
collapse and a cluster-pair collision would be the fundamental processes}.
Therefore in this paper, we would like to focus on such fundamental
dynamics disregarding the other non-gravity factors.

The cold collapsing process has been extensively studied 
as a crucial relaxation process of the collisionless systems 
such as elliptical galaxies where the stellar encounters are unimportant.
After the violent cold collapse, a steady state is generally formed, whose
density profile is well described by the de Vaucouleurs's $r^{1/4}$ law \cite
{deVaucouleurs48,Albada82,Jaffe87,Makino90,Aguilar90}.
The spherical averaged density profile is found to be $\rho \propto r^{-4}$ 
in spherical cold collapses\cite{Henon64,Albada82} 
and is found to be $\rho \propto r^{-3}-r^{-4}$ 
in the cluster-pair collision \cite{Merrall03}. 
The energy distribution is further discussed in \cite{Stiavelli85}. 
On the other hand, the velocity distribution has not 
yet been extensively studied except for 
the anisotropy of the velocity dispersion\cite{Binney87}. 
Recently, Merrall \textit{et al.} have numerically showed
that the radial velocity distribution in the central region of the bound
particles becomes Gaussian\cite{Merrall03}.
After the spherical collapse and the cluster-pair collision, 
Kanaeda and Morikawa showed that 
the velocity distribution of all bound particles becomes 
non-Gaussian\cite{Kanaeda03} which is described by the superposed-Gaussian.

From a viewpoint of statistical mechanics, we cannot naively expect the
ordinary Gaussian distributions for SGS, because the the long-range
interaction of gravity apparently \textit{violates additivity}, which is the
basic standpoint of the ordinary Boltzmann statistical mechanics\cite{Tsallis95}. 
Then the question is whether we can expect 
any universal distribution function for
SGS instead of the Gaussian distribution.

As one of the possible explanations for these non-Gaussian distributions in
the stationary state with large fluctuations of intensive quantities such as
temperature, Beck and Cohen\cite{Beck} proposed the superstatistics.
According to this proposal, statistical properties of the temperature
fluctuations determine overall non-Gaussian distributions. A special
choice of the fluctuation leads to the Tsallis statistics\cite{Tsallis}.

However, it is clear that the non-Gaussian properties of SGS are \textit{not
always observed} everywhere in the Universe. Moreover non-Gaussian
properties of SGS are quite diverse in general and we cannot expect the
completely universal properties of SGS. For example, we studied non-Gaussian
properties in the self-gravitating ring model\cite{Sota01}, where many
self-gravitating particles are constrained to move on a circular ring, which
is fixed in the three-dimensional space. 
Only at the intermediate energy scale where the specific heat becomes negative, 
only the \textit{halo} particles 
which belong to the intermediate energy scale, 
have shown non-Gaussian and power law velocity distribution: $f(v)\propto v^{-2}$. 
In this model, 
the existence of the \textit{halo} particles plays an essential role
in the appearance of non-Gaussian distributions. 
In our present paper therefore, 
we would like to clarify \textit{in which aspect of SGS and under
which conditions the universal non-Gaussian properties are observed}.

Non-Gaussian properties would be more naturally 
expected for the collisionless stage of the evolution ($t<t_{rel}$) 
than the later collisional stage of SGS ($t\gg t_{rel}$), 
where $t_{rel}$ is a local two-body relaxation time\cite{Spitzer} 
(In this paper, $t_{rel}$ is defined by Eq.(\ref{t_rel})). 
This is because in the former stage, 
the system is not thermally relaxed and the local equilibrium 
is not yet established. 
A long range dynamical correlation among the whole system 
develops though the long range interaction. 
This property would yield non-additivity of the system 
and manifest deviation from the ordinary Gaussian distributions. 
On the other hand, in the later stage, 
the local equilibrium is established through two-body encounters 
and the situation is similar to the ordinary statistical mechanics 
which admits Gaussian distributions. 
Thus we would like to concentrate on 
the \textit{collisionless stage of SGS }in this paper.

In section \ref{sec:VD}, 
we begin our study with typical simulations 
for spherical cold collapses and cluster-pair collisions. 
We find the same form of non-Gaussian velocity distribution in both cases, 
and then we explore four different models 
of the superposed-Gaussian distributions to describe 
this velocity distribution. 
In section \ref{sec:DT}, 
by analyzing the numerical data we show 
that the non-Gaussian velocity distribution observed in our simulation 
are well described by the ``Democratic Temperature(DT) distribution''. 
This DT distribution is consistent with the fact that 
we observe the linear relation 
between the temperature and the inner mass. 
In section \ref{sec:Universal}, we study the universality of this DT
distribution and show that the mixing property under the violent
gravitational process is the essence for the appearance of the DT
distribution. The effect of coherent motion in the velocity distribution is
discussed in section \ref{sec:unif}. The last section \ref{sec:con} is
devoted to the discussions and further developments of the present work.


\section{\label{sec:VD}the velocity distributions in N-body simulation}


A self-gravitating N-body system is described by the following Hamiltonian: 
\be
H=\sum_{i=1}^{N}\frac{\bm p_{i}^{2}}{2m}-\sum_{i<j}^{N}\frac{Gm^{2}}{\left\{
(\bm r_{i}-\bm r_{j})^{2}+\epsilon ^{2}\right\} ^{1/2}}, 
\label{H}
\ee
where $\bm r_{i}$ and $\bm p_{i}$ are respectively the position and the
momentum of the $i$th particle, $m$ is the mass of the each particle, and $G$
is the gravitational constant. For numerical simulations, we have to
introduce a cutoff parameter $\epsilon$. 
For the units of length, mass, and time, 
we use the initial system size $R$, the total mass $M:=Nm$, and the
initial free fall time $t_{ff}:=\sqrt{R^{3}/(GM)}$, respectively. 
The local two-body relaxation time $t_{rel}$ in this simulation 
can be written in the following form:
\be
t_{rel}=\frac{0.065\sigma^3(r)}{G^2\rho(r) m\ln(1/\epsilon)},
\label{t_rel}
\ee
where $\sigma(r)$ is a velocity dispersion and $\rho(r)$ is a mass density. 
We use a leap-frog symplectic integrator on GRAPE-5, a special-purpose computer
designed to accelerate N-body simulations\cite{GRAPE}. In all runs, the
numerical errors in the total energy $|\Delta E/E_{0}|$ have been controlled
to be less than $10^{-3}$.

In this section, we focus on the velocity distribution in the violent
gravitational process by N-body simulation. Especially we consider two
fundamental processes of violent gravitational dynamics; a spherical cold
collapse and a cluster-pair collision.


\subsection{\label{sec:SC}Spherical Cold collapse process}


We first glance at a typical example of a spherical cold collapse process
(run SC in Table.\ref{tab:initialSC}). All particles are homogeneously
distributed within a sphere of radius $R$. The system is composed of 
$N=5000$ particles and we set the vanishing virial ratio $|2K/W|=0$
initially, where $K$ and $W$ are the kinetic energy and the
potential energy of the whole system, respectively.

\begin{figure}[ht]
\includegraphics[width=8cm]{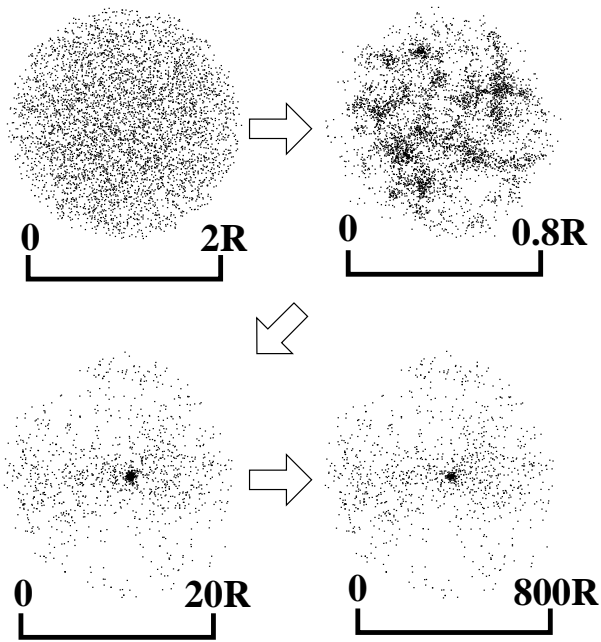}
\caption{A snapshot of a spherical cold collapse case 
(run SC in Table \ref{tab:initialSC}): 
top left ($t=0$), top right ($t=1t_{ff}$), bottom left ($t=5t_{ff}$), 
and bottom right ($t=100t_{ff}$).}
\label{SC-snapshot}
\end{figure}

Fig.\ref{SC-snapshot} shows snapshots of particle distributions at different
times. Most particles rapidly collapse into the center within the free-fall
time $t\sim t_{ff}$. After this collapse, some particles obtain positive
energy and escape from the system, and the rest particles remain bounded and
gradually expand leaving a tight core at the center. This is a typical
process of the formation of the core-halo structure for SGS.

\begin{figure}[th]
\includegraphics[width=8cm]{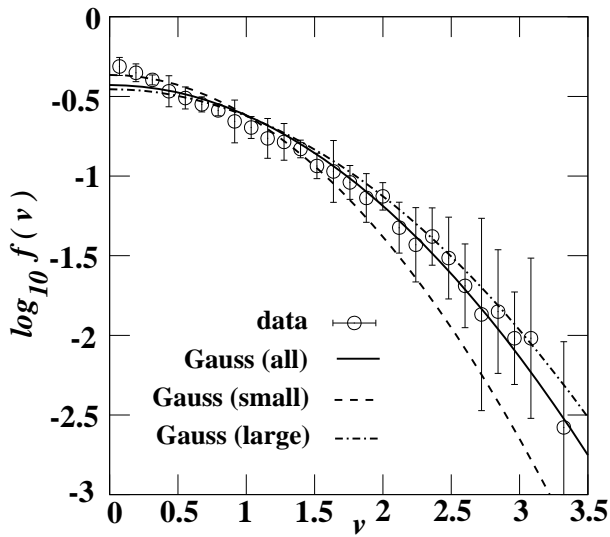}
\caption{A velocity distribution of a spherical cold collapse 
(run SC in Table \ref{tab:initialSC}) at $t=10t_{ff}$. 
Three best-fitted Gaussian distributions are superposed. 
Gauss (all) is fitted by use of the all data. 
Gauss (small) and Gauss (large) are fitted by use of data 
included for only small velocity and only large velocity, respectively.}
\label{SCv}
\end{figure}

We now focus on the velocity distributions of the particles. 
Since we are interested in the global property, 
we extract the one-dimensional velocity distribution function 
combining all-directional components of velocity distributions.
We will examine the anisotropy in the velocity distribution 
in Sec.\ref{sec:unif}. 
The velocity distribution thus obtained 
is shown for bound particles, which have negative energy, in Fig.\ref{SCv}. 
This velocity distribution
is apparently different from the ordinary Gaussian distribution, especially
in the small velocity region. There is an apparent cusp at the center. 
This velocity distribution has been
stably observed just after the collapse during the whole
simulation time ($t\sim t_{ff}\rightarrow 1000t_{ff}$). We emphasize that
this non-Gaussian velocity distribution is quite universal and robust in
various cold collapsing process, as we will demonstrate briefly.

\subsection{\label{sec:CC}Cluster-pair collision process}


We turn our attention to another case of 
the violent gravitational process: 
The cluster-pair collision. 
Because we would like to clarify the basic process, 
we choose the simplest head-on collision 
(run CC in Table.\ref{tab:initialCC}). 
Each cluster has an equal number of particles and 
all particles are homogeneously distributed 
in each sphere of radius $R$. 
The initial velocity distribution is set to be Gaussian 
and the initial virial ratio is $1$. 
We set the initial separation of the pair to $6R$ 
along the $x$ axis.

\begin{figure}[tbp]
\includegraphics[width=8cm]{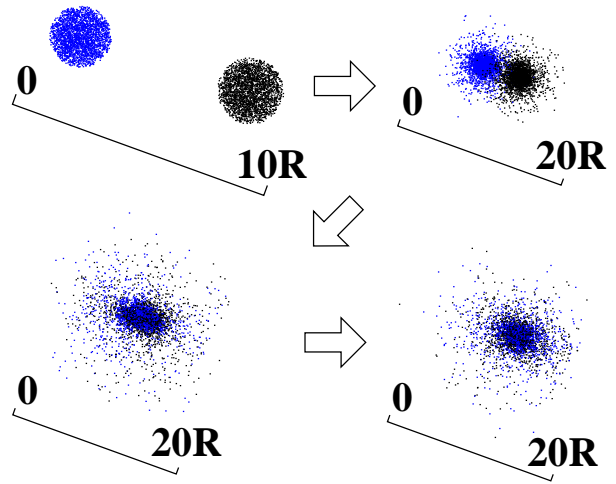}
\caption{A snapshot of a cluster collision model 
(run CC in Table \ref{tab:initialCC}): 
top left ($t=0$), top right ($t=5t_{ff}$), bottom left ($t=20t_{ff}$), 
and bottom right ($t=500t_{ff}$).}
\label{CC-snapshot}
\end{figure}

\begin{figure}[ht]
\includegraphics[width=8cm]{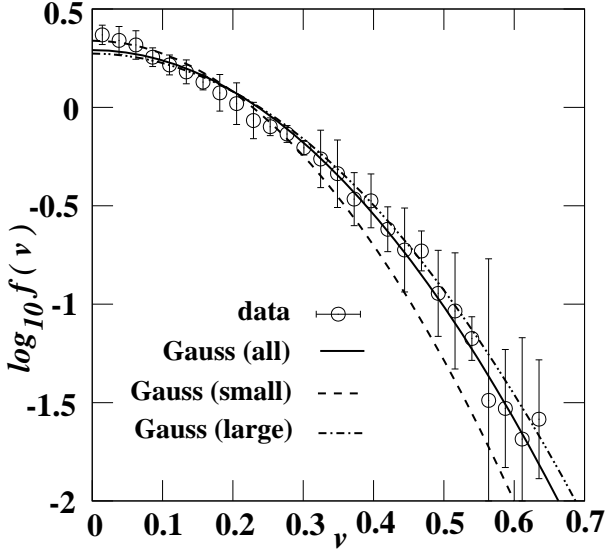}
\caption{Same as Fig.\ref{SCv}, 
but for a cluster-pair collision model 
(run CC in Table \ref{tab:initialCC}) at $t=500t_{ff}$.}
\label{CCv}
\end{figure}

Fig.\ref{CC-snapshot} shows the snapshots of particle distributions for the
run CC at different times. The cluster-pair collides at time $t\sim 20t_{ff}$
and then gradually merges into a single cluster. The velocity distribution
profile after this merging is shown in Fig.\ref{CCv}. The velocity
distribution profile is very similar to the previous spherical collapse
case(run SC). The excess of velocity distribution at the small value is
prominent as before.

In both cases, 
we notice that the temperature or the velocity dispersion 
is different from place to place; 
the center of the core is much hotter than the outskirts of the system. 
Therefore it is clear from the beginning that
the obtained velocity distribution cannot be fitted by a single Gaussian
distribution. Thus we propose to describe these systems by the superposition
of Gaussian distributions with various temperatures. This consideration is
very natural because the collisionless SGS allows the coexistence of many
temperatures. This is because (a) the collisionless SGS is not yet
thermally relaxed, and (b) the virial relation implies that the system shows
negative specific heat, which spontaneously yields inhomogeneous temperature.

In general, superposition of Gaussian velocity distribution $f_{SG}(v_{i})$
has the form: 
\begin{eqnarray}
f_{SG}(v_{i}) &=&A\int_{0}^{M}dM^{\prime }\frac{1}{\sqrt{2\pi T(M^{\prime })}%
}e^{-v_{i}^{2}/2T(M^{\prime })},  \nonumber \\
&=&A\int_{0}^{T}dT^{\prime }\left| \frac{dM}{dT^{\prime }}\right| \frac{1}{%
\sqrt{2\pi T^{\prime }}}e^{-v_{i}^{2}/2T^{\prime }},  \label{SGauss}
\end{eqnarray}
where $A$ is a normalization constant. The parameter function $M$ can be any
function of temperature. However now in the present spherical symmetric
case, we can choose it to be the inner mass: 
$M_{r}:=\int_{0}^{r}\rho \left( r^{\prime }\right) 4\pi r^{\prime 2}dr^{\prime }$. 
This choice is not at all trivial and will be demonstrated briefly. 
Any special choice of the weight function for the superposition yields 
in general arbitrary distribution functions. 
In order to avoid such arbitrariness, 
we will choose the simplest weight functions. 
Here we consider four natural models of the superposed-Gaussian distribution.

\begin{itemize}
\item  Model 1 (GT) Gaussian-weighted superposition of Gaussian
distributions with various temperatures.

The temperature distribution is the Gaussian with the dispersion $\tau $,
and GT distribution $f_{GT}(v_{i},\tau )$ takes the following form:

\begin{eqnarray}
f_{GT}(v_{i},\tau ) &=&\int_{0}^{\infty }dT\frac{1}{\sqrt{2\pi \tau }}\exp \left[
-\frac{T^{2}}{2\tau }\right] \frac{1}{\sqrt{2\pi T}}\exp \left[ -\frac{%
v_{i}^{2}}{2T}\right]  \nonumber \\
&=&\frac{1}{2\pi \sqrt{\tau }}\left[ \frac{\tau ^{1/4}\Gamma
(1/4){}_{0}F_{2}(1/2,3/4;-v_{i}^{4}/(32\pi ))}{2^{3/4}}\right.  \nonumber \\
&&\left. -\sqrt{2\pi }|v_{i}|{}_{0}F_{2}(3/4,5/4;-v_{i}^{4}/(32\tau ))\right.
\nonumber \\
&&\left. +\frac{v_{i}^{2}\Gamma (3/4){}_{0}F_{2}(5/4,3/2;-v_{i}^{4}/(32\pi ))%
}{2^{1/4}\tau ^{1/4}}\right] ,  \label{GT}
\end{eqnarray}
where $\Gamma (x)$ is a Gamma function and ${}_{p}F_{q}(\alpha _{1},\dots
,\alpha _{p};\beta _{1},\dots ,\beta _{q};z)$ is the generalized
Hypergeometric function.

\item  Model 2 (DT) Equal-weighted superposition of Gaussian distributions
with various temperatures.

This case corresponds to $dT/dM=const.$ in terms of the parameter function $%
M $, and we call the velocity distribution in this case as ``democratic
temperature distribution''$f_{DT}(v_{i})$:

\begin{eqnarray}
f_{DT}(v_{i}) &=&\frac{1}{T}\int_{0}^{T}{dT^{\prime }\frac{1}{\sqrt{2\pi
T^{\prime }}}e^{-v_{i}^{2}/\left( 2T^{\prime }\right) }},  \nonumber \\
&=&\frac{1}{T}\left[ {\sqrt{\frac{2T}{\pi }}e^{-v_{i}^{2}/(2T)}-\left| {v_{i}%
}\right| \left\{ {1-\mbox{Erf}\left( {\frac{\left| {v_{i}}\right| }{\sqrt{2T}%
}}\right) }\right\} }\right],\nonumber\\
\label{DT}
\end{eqnarray}
where $\mbox{Erf}(x)$ is the error function: 
\[
\mbox{Erf}(x):=\frac{2}{\sqrt{\pi }}\int_{0}^{x}dte^{-t^{2}}. 
\]

\item  Model 3 (G$\sigma $ ) Gaussian-weighted superposition of Gaussian
distributions with various standard deviations $\sigma:=\sqrt{T}$.

The distribution of the standard deviation is the Gaussian with the
dispersion $\gamma^2$, and G$\sigma $ distribution 
$f_{G\sigma }(v_{i},\gamma )$ takes the following form:

\begin{eqnarray}
f_{G\sigma }(v_{i},\gamma ) &=&\int_{0}^{\infty }d\sigma \frac{1}{\sqrt{2\pi 
}\gamma }\exp \left[ -\frac{\sigma ^{2}}{2\gamma ^{2}}\right] \frac{1}{\sqrt{%
2\pi }\sigma }\exp \left[ -\frac{v_{i}^{2}}{2\sigma ^{2}}\right] ,  \nonumber
\\
&=&\frac{K_{0}\left( \left| {v_{i}/\gamma }\right| \right) }{2\pi \gamma },
\label{Gsigma}
\end{eqnarray}
where $K_{0}(z)$ is the modified Bessel function.

\item  Model 4 (D$\sigma $ ) Equal-weighted superposition of Gaussian
distributions with various standard deviation $\sigma $.

Since $d\sigma /dM=const.$, D$\sigma $ distribution 
$f_{D\sigma }(v_{i})$ has the following form:

\begin{eqnarray}
f_{D\sigma }(v_{i}) &=&\frac{1}{\sigma }\int_{0}^{\sigma }d\sigma ^{\prime }%
\frac{1}{\sqrt{2\pi }\sigma ^{\prime }}e^{-v_{i}^{2}/\left( 2\sigma ^{\prime
2}\right) },  \nonumber \\
&=&\frac{\Gamma (0,v_{i}^{2}/\left( {2\sigma ^{2}}\right) )}{2\sqrt{2\pi }%
\sigma },  \label{Dsigma}
\end{eqnarray}
where $\Gamma (z,p)$ is the incomplete Gamma function.
\end{itemize}

In the next section, we examine the above four models in relation with our
non-Gaussian velocity distributions.


\section{\label{sec:DT}Democratic Temperature (DT) distribution}


As shown in the preceding section, 
the velocity distribution functions for both processes, 
spherical cold collapses and cluster-pair collisions, 
take the same non-Gaussian form. 
Using the most natural four models of superposed-Gaussian distributions, 
we try to fit our velocity distribution function and
to extract any universal character of SGS.

\begin{figure}[ht]
\includegraphics[width=8cm]{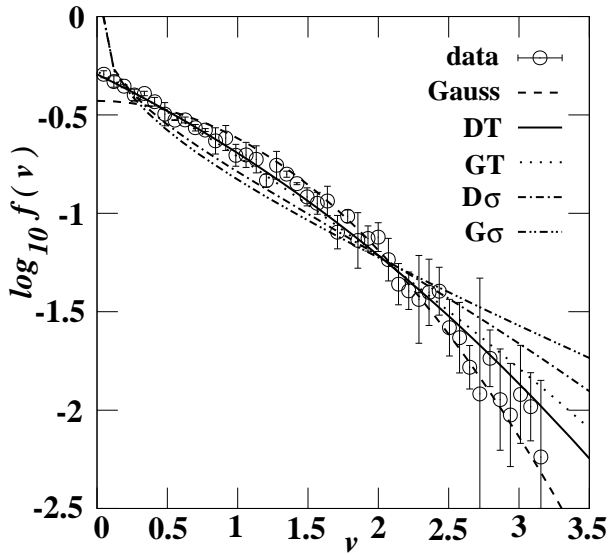}
\caption{A linear-log plot for velocity distribution for the case of a
spherical cold collapse (run SC) at $t=10t_{ff}$. 
The best fit Gaussian and the four superposed-Gaussian model 
(Eqs.(\ref{GT}),(\ref{DT}),(\ref{Gsigma}), and (\ref{Dsigma})) 
with a best fit parameter are superposed. 
The $\protect\chi^2$ of each model is 
0.00095(Gauss), 0.00037(DT), 0.00048(GT), 
0.0016(D$\protect\sigma$), and 0.0025(G$\protect\sigma$). 
The best fit model is a DT distribution. }
\label{SCv-fit}
\end{figure}

\begin{figure}[ht]
\includegraphics[width=8cm]{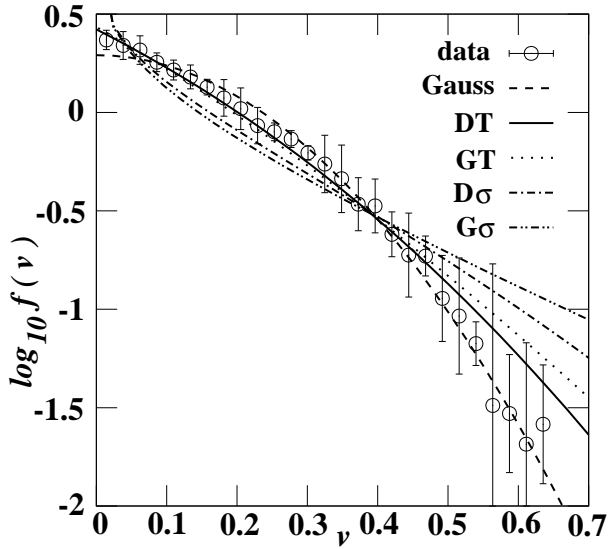}
\caption{Same as Fig.\ref{SCv-fit}, but for a cluster-pair collision
model (run CC) at $t=500t_{ff}$. 
The $\protect\chi^2$ of each model is
0.0030(Gauss), 0.0021(DT), 0.0030(GT), 0.010(D$\protect\sigma$), 
and 0.015(G$\protect\sigma$). The best fit model is a DT distribution. }
\label{CCv-fit}
\end{figure}

The fitting results are shown in Fig.\ref{SCv-fit} and \ref{CCv-fit}.
Using the $\chi ^{2}$ analysis, the best fit model turns out to be the
DT distribution in both cases.

In order to investigate the structure of velocity space for bound particles
in detail, we divide the whole particles into several shells with equal
number of particles. We introduce the inner mass coordinate $M_{r}:=4\pi
\int_{0}^{r}dr^{\prime }r^{\prime 2}\rho (r^{\prime })$ and consider the
averaged quantities within each shell as local variables.

\begin{figure}[ht]
\includegraphics[width=8cm]{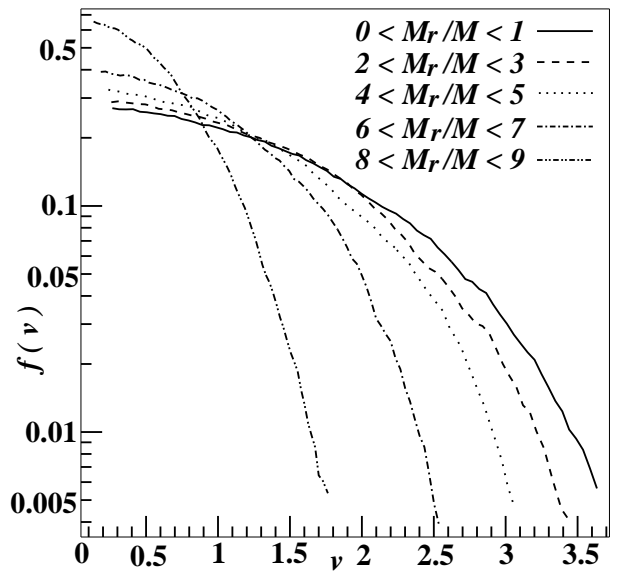}
\caption{A linear-log plot for a velocity distribution of each shell for the
case of a spherical cold collapse (run SC). We divided the whole bound
particles into 10 shells including an equal mass and calculated the velocity
distribution by use of the data from $t=10t_{ff}$ and $t=20t_{ff}$. 
}
\label{SCv-shell}
\end{figure}
\begin{figure}[ht]
\includegraphics[width=8cm]{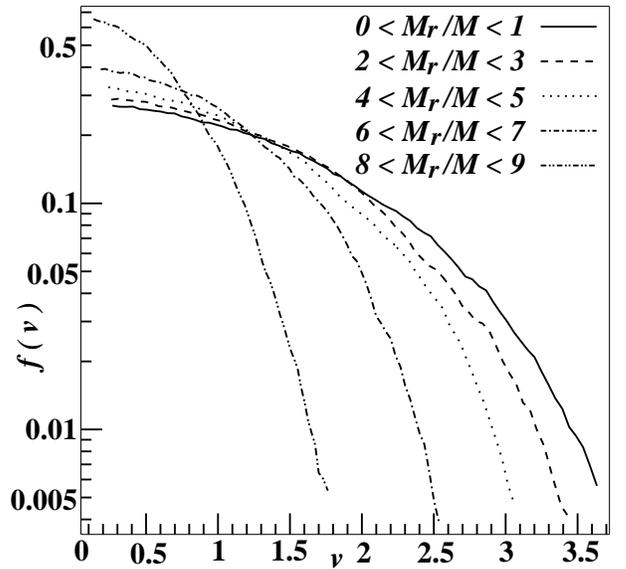}
\caption{Same as Fig.\ref{SCv-shell}, 
but for a cluster-pair collision (run CC). 
We calculated the velocity distribution 
by use of the data from $t=500t_{ff}$ and $t=510t_{ff}$.}
\label{CCv-shell}
\end{figure}

The velocity distribution of each shell shows almost Gaussian 
(Fig.\ref{SCv-shell} and \ref{CCv-shell}). 
These figures show that each shell has a different temperature and 
suggest that the previous non-Gaussian distribution can be described 
by the superposition of Gaussians with various temperatures. 
The temperature is the highest at the most inner shell and
monotonically decreases toward outside shells.

\begin{figure}[ht]
\includegraphics[width=8cm]{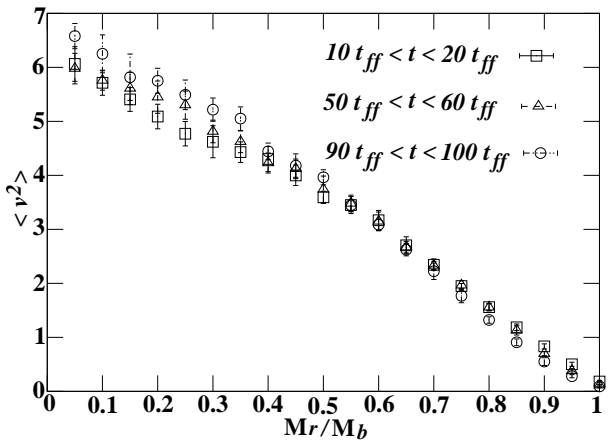}
\caption{A mass dependence of velocity dispersion ($<v^2(M_r)>$) for the
case of a spherical cold collapse (run SC). The mass is normalized by the
mass of whole bound particles $M_b$. The velocity dispersion is 
time averaged and the error bar is calculated from the time fluctuation. }
\label{SCv2}
\end{figure}
\begin{figure}[ht]
\includegraphics[width=8cm]{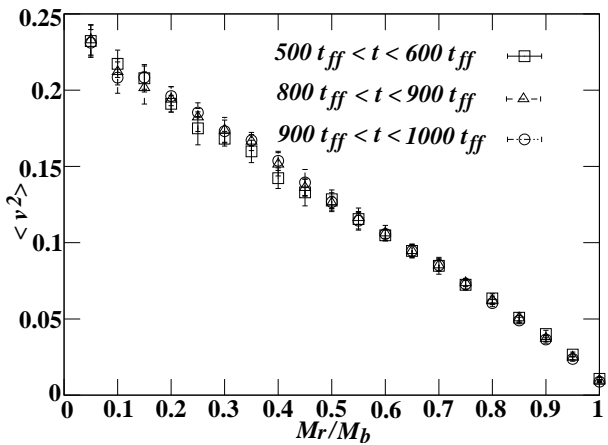}
\caption{Same as Fig.\ref{SCv2}, 
but for a cluster-pair collision (run CC).}
\label{CCv2}
\end{figure}

The local velocity dispersion $<v^{2}>$ is plotted 
against the inner mass $M_{r}$ in Fig.\ref{SCv2} and \ref{CCv2}. 
In both cases, it is remarkable
that the velocity dispersion decreases linearly in the inner mass $M_{r}$.
We emphasize here 
that this linear relation is quite robust and universal.
Actually this relation is observed in our two processes 
with wide range class of initial conditions, 
which will be further discussed in the subsequent sections. If we
carefully look at Fig.\ref{SCv2}, the temperature gradient in the inner
region gradually reduces in time, while it remains constant in the outer
region. Since the mass density at the inner region increases in time and
eventually the effect of two-body relaxation turns on, 
such a tilt of the temperature gradient is 
thought to be caused by this collisional effect. 
A careful check of our numerical simulation supports this interpretation.
Therefore the linearity of the temperature gradient is 
thought to be the property of the collisionless SGS. 
Further collisional effect would eventually yield 
the uniform temperature distributions.

The almost Gaussian velocity distribution in each shell 
guarantees the form of Eq.(\ref{SGauss})
and the linearity of the temperature-mass relation leads to 
the relation $|dM/dT|=const.$ in Eq.(\ref{SGauss}), 
which indicates the DT distribution for the velocity distributions. 
This is perfectly consistent with the result that 
the DT distribution is the best fit to the velocity distribution 
among several models which we examined.

\begin{figure}[ht]
\includegraphics[width=8cm]{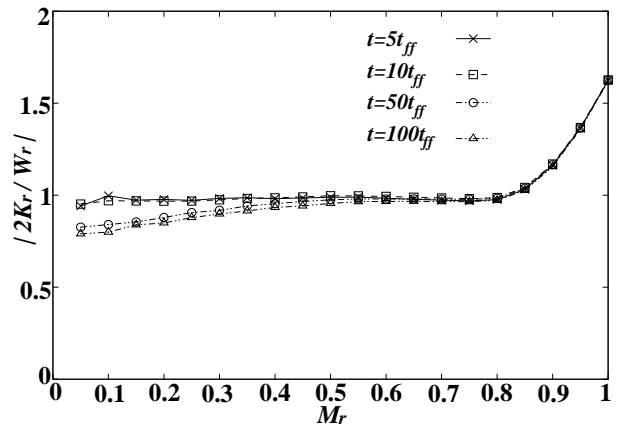}
\caption{The local virial relation for run SC. The ratio $\left|2\overline{K}%
_{r}/\overline{W}_{r}\right|$ is plotted as a function of $M_{r}$. The
virial ratios of each shell at different times 
($t=5t_{ff}$, $10t_{ff}$, $50t_{ff}$, and $100t_{ff}$) are superposed.}
\label{SC-virial}
\end{figure}

\begin{figure}[ht]
\includegraphics[width=8cm]{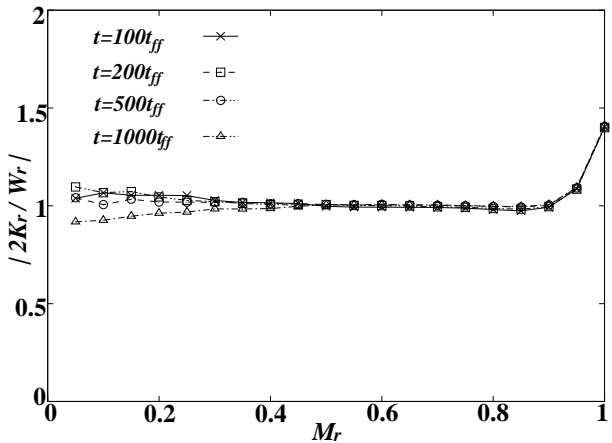}
\caption{Same as Fig.\ref{SC-virial}, 
but for a cluster-pair collision (run CC). 
The virial ratios of each shell at different times 
($t=100t_{ff}$, $200t_{ff}$, $500t_{ff}$, and $1000t_{ff}$)
are superposed. }
\label{CC-virial}
\end{figure}

Besides this linearity in the diagram of 
the local velocity dispersion $<v^{2}>$ and the inner mass $M_{r}$, 
we observe another prominent fact that
the virial relation between the potential energy and kinetic energy holds 
\textit{locally} at each spatial region. 
Using the local kinetic energy $K_{r}$ 
and the local potential energy $W_{r}$ inside the radius $r$, 
we can locally define the virial ratio $|2K_{r}/W_{r}|$. 
The time evolution of this local virial ratio is depicted 
in Fig.\ref{SC-virial} and \ref{CC-virial}. 
The value of the local virial ratio stays almost unity 
within the error $\pm 10\%$ everywhere anytime. 
This suggests that not only the whole system
but also each shell is locally virialized. 
This fact may be deeply related with 
the robustness of the DT distribution, 
and will be separately discussed in \cite{Sota04}.


\begin{figure}[ht]
\includegraphics[width=8cm]{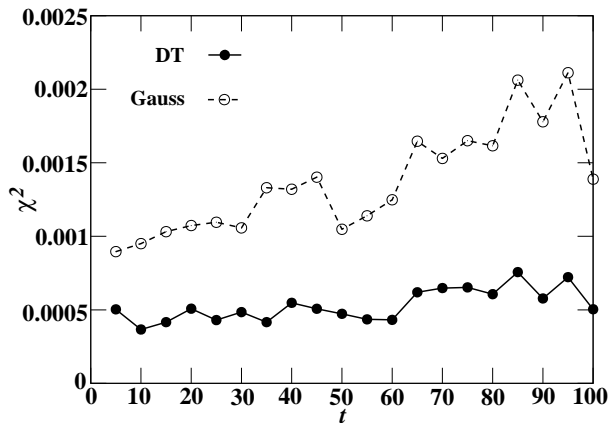}
\caption{Time evolution of $\protect\chi^2_{DT}$ of DT distribution and $%
\protect\chi^2_{Gauss}$ of Gaussian for the case of a spherical cold
collapse (run SC). }
\label{SCv-chi2}
\end{figure}

\begin{figure}[ht]
\includegraphics[width=8cm]{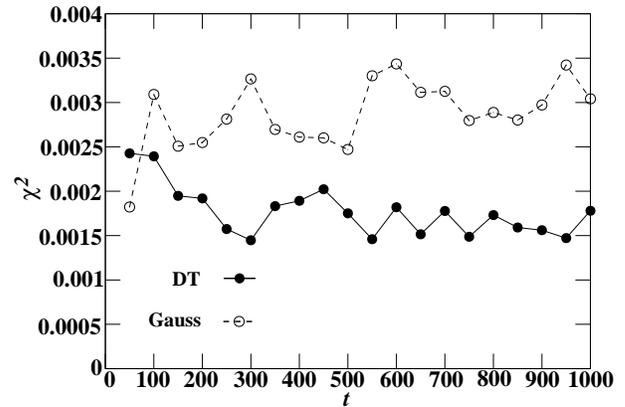}
\caption{Same as Fig.\ref{SCv-chi2}, 
but for a cluster-pair collision (run CC).}
\label{CCv-chi2}
\end{figure}

In order to examine the stability of DT distribution, we calculate the time
evolution of the $\chi ^{2}$ value for DT ($\chi _{DT}^{2}$) and Gaussian
distributions ($\chi _{Gauss}^{2}$), 
which is depicted in Fig.\ref{SCv-chi2} and \ref{CCv-chi2}. 
Immediately after the collapse or the merging, 
the velocity distribution always becomes the DT distribution 
during the whole period of our simulations. 
This robustness makes DT distribution 
one of the most characteristic properties of collisionless SGS.



\section{\label{sec:Universal}Universality of DT distribution - degree of
violence}


As we have seen in the preceding section, DT distribution, 
supported from the linear $<v^{2}>$ - $M_{r}$ relation, 
is one of the most relevant characteristics of SGS 
such as the spherical cold collapses and cluster-pair collisions. 
We would like to further clarify the condition 
for the appearance of the DT distribution function. 
From the beginning of our numerical experiments, 
we have implicitly chosen the violent processes 
such as cold collapse and the quiet cluster-pair collision. 
We had in mind the expectation that 
the strong mixing property associated with 
the violent gravitational process 
would yield the most prominent characteristics 
for the collisionless SGS. 
Actually it is well known that 
a violent mixing becomes an important factor to 
realize the quasi-equilibrium virialized state\cite{Lynden67}.

In this section, 
we will quantitatively demonstrate this expectation.
Especially we would like to explore the correlation 
between the DT distribution and the degree of violence of the processes.

We control the degree of violence of the process by the choice of initial
conditions in our numerical experiments. The initial conditions of all runs
are listed in Table.\ref{tab:initialSC} 
for spherical collapses and Table.\ref{tab:initialCC} for cluster-pair collisions.

\begin{table}[tbp]
\caption{Initial condition for spherical collapse case. $N$ is a number of
particles and $|2K/W|$ is an initial virial ratio and $\protect\rho\propto
r^{-a}$ is an initial density profile and $\protect\epsilon$ is a cutoff
parameter in Eq.(\ref{H}). }
\label{tab:initialSC}%
\begin{ruledtabular}
\begin{tabular}{lrrrr}
run
& N
& $|2K/W|$ 
& $\rho\propto r^{-a}$
& $\epsilon$ 
\\\hline
SC  & 5000   & 0   &  0   & $2^{-8}$ \\
SCN1 & 10000  & 0   &  0   & $2^{-8}$ \\
SCN2 & 50000  & 0   &  0   & $2^{-8}$ \\
SCV1  & 5000  & 0.1 &  0   & $2^{-8}$ \\
SCV2  & 5000  & 0.2 &  0   & $2^{-8}$ \\
SCV3  & 5000  & 0.3 &  0   & $2^{-8}$ \\
SCV4  & 5000  & 0.4 &  0   & $2^{-8}$ \\
SCV5  & 5000  & 0.5 &  0   & $2^{-8}$ \\
SCV6  & 5000  & 1.0 &  0   & $2^{-8}$ \\
SCA1  & 5000  & 0   &  0.5 & $2^{-8}$ \\
SCA2  & 5000  & 0   &  1.0 & $2^{-8}$ \\
SCA3  & 5000  & 0   &  1.5 & $2^{-8}$ \\
SCA4  & 5000  & 0   &  2.0 & $2^{-8}$ \\
SCC1  & 5000  & 0   &  0   & $2^{-4}$ \\
SCC2  & 5000  & 0   &  0   & $2^{-6}$ \\
SCC3  & 5000  & 0   &  0   & $2^{-10}$ \\
SCS1 & 5000   & 0.1\footnotemark[1]& 0 & $2^{-8}$ \\
SCS2 & 5000   & 0.5\footnotemark[1]& 0 & $2^{-8}$ \\
SCS3 & 5000   & 1.0\footnotemark[1]& 0 & $2^{-8}$ \\
\end{tabular}
\end{ruledtabular}
\footnotetext[1]{The kinetic energy is contributed by only the rigid
 rotation around the $z$ axis.}
\end{table}
%

\begin{table}[tbp]
\caption{Initial condition for cluster collision case. $N$ is the total number
of particles and the number of particles in each cluster is $N/2$. $|2K/W|$ is
an initial virial ratio of each cluster where the kinetic energy $K$ is
contributed by only the random motion. $K_{rot}$ is a kinetic energy 
by only the rotation with an orbital angular moment around the z axis 
($L_z$) initially.}
\label{tab:initialCC}%
\begin{ruledtabular}
\begin{tabular}{lrrr}
run
& $N$
& $|2K/W|$
& $|2K_{rot}/W|(L_z)$ \\\hline
CC  & 5000   & 1  &  0  \\
CCL1  & 5000   & 1  &  0.1  \\
CCL2  & 5000   & 1  &  0.2  \\
\end{tabular}
\end{ruledtabular}
\end{table}
%

As the indicator for the degree of violence of the process, we use the total
energy fluctuation of particles; more violent the process the more rapid the
energy mixing of each particles. More precisely, the energy fluctuation of
the $i$th particle is defined as 
\begin{equation}
\sigma _{e;i}:= \sqrt{\overline{e_{i}^{2}(t)}-\overline{e_{i}(t)}^{2}},
\end{equation}
where $e_{i}(t)$ is the energy of $i$th particle at time $t$ and 
$\overline{\ast }$ is the time averaged value of $\ast $ during the
whole simulation period. Then we define the total energy fluctuation of the
system as the sum of $\sigma _{e;i}$ for all particles as 
\begin{equation}
\sqrt{\Delta E^{2}}:= \sum_{i=1}^{N}\sigma _{e;i}.
\end{equation}

In order to check the efficiency of $\sqrt{\Delta E^{2}}$ 
as the indicator of the degree of violence, 
we calculate the correlations with other physical quantities 
which are naturally expected to be connected 
with the strength of the mixing. 
When violent mixing in the phase space occurs, 
some of the particles obtain enough energy to escape from the system 
through the energy exchange by the potential oscillation. 
Then the rest of the particles becomes more tightly bounded 
due to the extraction of energy by the escaping particles. 
Therefore both (a) the system size and 
(b) the total energy of the bound particles, $E_{b}(<0)$, 
will become smaller after the collapse,
provided the mixing is strong and effective. 
As the quantitative indicator for (a), 
i.e., the size of the system after the collapse, 
we introduce the half-mass radius 
divided by it's initial value: $R_{hm}$. 
Further, we need to take ensemble average $<\ast >$ 
of all those quantities for the regular continuous indicator for the process.

\begin{figure}[ht]
\includegraphics[width=8cm]{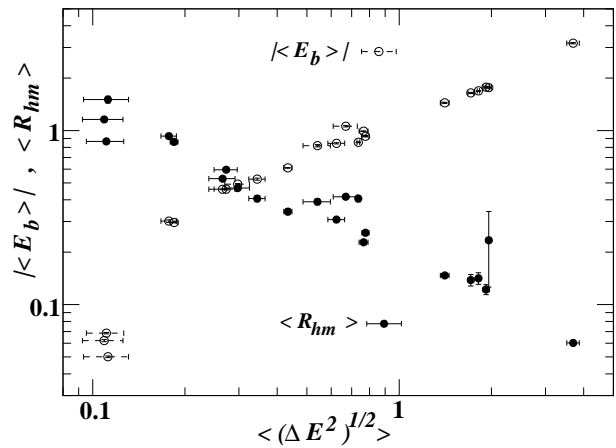}
\caption{The half-mass radius normalized by it's initial value at $t=0$, 
$R_{hm}$ and the absolute value of the total energy of bound particles, 
$|E_b| $ are plotted against $\protect\sqrt{\Delta E^2} $ for all cases in
Table.\ref{tab:initialSC} and Table.\ref{tab:initialCC}. We take the
ensemble average for each data with several times.
It ranges from $t=10t_{ff}$ to $t=100t_{ff}$ with $10t_{ff}$ intervals 
in the spherical case, and from $t=100t_{ff}$ and $t=1000t_{ff}$ with $100t_{ff}$
intervals in cluster collisions. 
The error bar is calculated from the ensemble average. }
\label{hmr_edisp}
\end{figure}

\begin{figure}[ht]
\includegraphics[width=8cm]{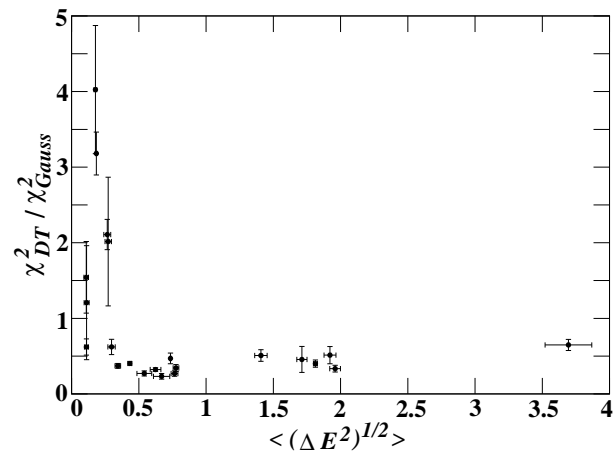}
\caption{$\protect\chi^2_{DT}/\protect\chi^2_{Gauss}$ 
for all cases in Table.\ref{tab:initialSC} 
and Table.\ref{tab:initialCC}. 
$\protect\chi^2$ is a time averaged 
from $t=10t_{ff}$ and $t={100t_{ff}}$ at $10t_{ff} $ intervals
for the spherical collapse case, 
from $t=100t_{ff}$ and $t={1000t_{ff}}$ at $100t_{ff}$ intervals 
for the cluster collision case. 
The error bar is calculated from the time fluctuation. }
\label{dt_gauss}
\end{figure}

\begin{figure}[ht]
\includegraphics[width=8cm]{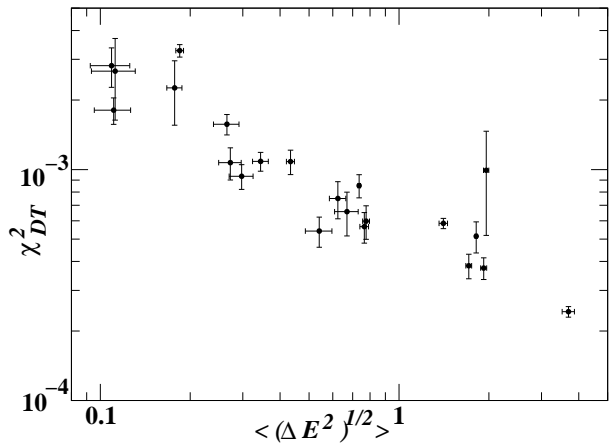}
\caption{Same as Fig.\ref{dt_gauss}, but for $\protect\chi^2_{DT}$.}
\label{edisp_chi2}
\end{figure}

As in shown in Fig.\ref{hmr_edisp}, $<\sqrt{\Delta E^{2}}>$ is apparently
correlated with both $<R_{hm}>$ and $<E_{b}>$. This fact supports that the
total energy fluctuation $<\sqrt{\Delta E^{2}}>$ is actually an 
effective indicator of the degree of strength of the mixing.

Keeping in mind that $<\sqrt{\Delta E^{2}}>$ is 
a good indicator of the violence of the process, 
we now examine the correlation between the degree of violence 
in the initial stage and the occurrence of the DT distribution
for bound particles in the quasi-equilibrium stage. 
We calculated $\chi^{2}$ values for the fitting 
of velocity distributions with DT distribution and
Gaussian distribution for all runs. 
The result is shown in Fig.\ref{dt_gauss}, 
where the ratio of $\chi _{DT}^{2}$ of DT distribution to 
$\chi_{Gauss}^{2}$ of Gaussian distribution is plotted against 
$<\sqrt{\Delta E^{2}}>$ for all initial conditions. 
The value is greater than one for the simulations 
with small value of $<\sqrt{\Delta E^{2}}>$, 
which means that the velocity distribution is 
better fitted by Gaussian rather than DT distribution. 
On the other hand, 
for the larger values of $<\sqrt{\Delta E^{2}}>$, 
the data is better fitted by DT distribution rather than Gaussian. 
Thus DT velocity distribution is always favorable 
in the simulation with violent processes.

This point is further clarified in Fig.\ref{edisp_chi2}, 
where we plotted the value of $\chi_{DT}^{2}$ 
against $<\sqrt{\Delta E^{2}}>$. 
It is apparent that the $\chi_{DT}^{2}$ monotonically decreases 
for increasing $\sqrt{\Delta E^{2}}$. 
These results manifestly support our expectation 
that the DT velocity distribution is associated 
with the strength of the violent mixing.

From these results, we can claim that the DT velocity distribution function
is universal for various initial conditions provided that the violent
gravitational process is included. Moreover, the DT distribution becomes more
prominent for stronger violent processes.


\section{\label{sec:unif}Universality of DT distribution - degree of
coherence}


We have so far studied the universality of DT distribution and its
correlation with the degree of violence of the dynamical process. 
Here in this section, 
we further discuss another aspect of universality of DT distribution function.

In the above argument, we have neglected possible anisotropy in the velocity
space; we have extracted one dimensional velocity data set by simply
combining all the velocity components. However actually in the processes of
spherical collapse and cluster-pair collision, the velocity space is
anisotropic in general. Further the anisotropy in the rotating collapse
(runs SCS1-3 in Table\ref{tab:initialSC}) is apparent.

\begin{figure}[th]
\includegraphics[width=8cm]{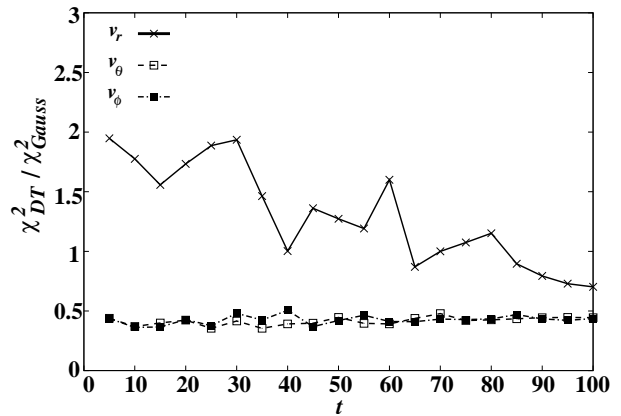}
\caption{Time evolution of $\protect\chi_{DT}^{2}$ of DT distribution
divided by the $\protect\chi_{Gauss}^{2}$ for the case of a spherical cold
collapse (run SC). }
\label{SCv-chi2-comp}
\end{figure}

For example, we first consider the spherical cold collapse (run SC). 
Fig.\ref{SCv-chi2-comp} shows the measure of validity of DT distribution, 
i.e., the time evolution of $\chi _{DT}^{2}$ divided by $\chi_{Gauss}^{2}$. 
The radial velocity distribution is better fitted by Gaussian for a while 
until $t\sim 50t_{ff}$ and after that time DT distribution becomes better. 
On the other hand, 
the tangential velocity distribution is always well fitted 
by DT distribution during all our simulation period.

This behavior will be related with the coherent motion of the particles.
More precisely, 
in the early stage of the evolution, 
there exists a coherent radial motion 
such as a collapse and a subsequent bounce, 
which eventually decays. 
This coherent motion does not affect the tangential velocity distribution. 
Moreover this coherent motion may inherit 
some amounts of randomness in the initial distribution of particles. 
Therefore in the early stage, 
only the radial velocity component is dominated 
by the coherent motion with this initial spatially random distribution, 
and eventually recovers the intrinsic DT distribution 
when the coherent motion decays.

\begin{figure}[ht]
\includegraphics[width=8cm]{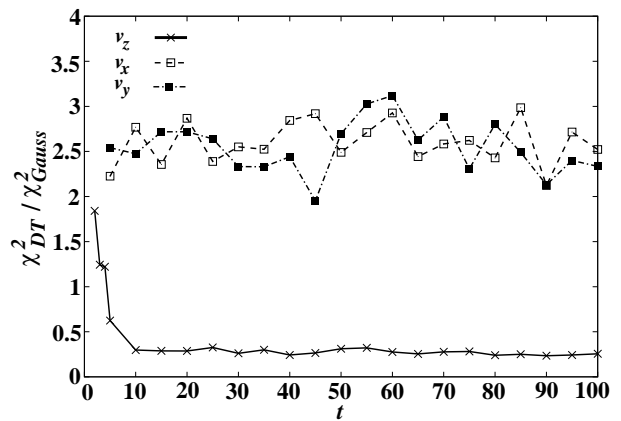}
\caption{Same as Fig.\ref{SCv-chi2-comp}, 
but for the case of a spherical collapse 
with the angular momentum $L_z$ (run SCS2).}
\label{SCv2-comp}
\end{figure}

The above consideration is also applicable for the anisotropy in the
rotating collapse (runs SCS1-3 in Table\ref{tab:initialSC}). 
Fig.\ref{SCv2-comp} shows
the measure of validity of DT distribution, 
i.e., the time evolution of $\chi_{DT}^{2}$ 
divided by $\chi_{gauss}^{2}$ in this case. 
Only the $v_{z}$-component shows DT distribution all the time, 
and the $v_{x},v_{y}$-components, the direction of coherent rotation, 
never show DT distribution during our simulation time.

Therefore the condition for the appearance of the universal DT distribution
would be, besides the violence of the system, the decay of the coherent
motion, which keeps the information of initial conditions.

\section{\label{sec:con}Conclusions and Discussions}


We have studied the velocity distribution function of 
self-gravitating systems (SGS) 
which experienced violent gravitational processes 
by use of N-body simulation method. 
Especially we have chosen two fundamental processes 
which mimic the galaxy formation dynamics; 
the spherical cold collapse and the cluster-pair collision.

In both cases, after the collapses or the collisions, 
all the bound particles form a single stationary state, 
which are characterized by the local virial relation 
and the linearity in the mass-temperature relation. 
In this stationary state, 
the velocity distribution is well described by 
\textit{the democratic (=equally weighted) superposition of Gaussian distributions of
various temperatures (DT distribution)}.

This DT velocity distribution is robust against various initial conditions
such as the change of the particle number, the virial ratio, 
and the density profile. 
Moreover, using the half-mass radius and the root mean square of energy variation 
as a measure of the degree of violent mixing, 
we found a firm positive correlation 
between the DT distribution and the degree of violent mixing. 
This correlation suggests that \textit{the DT distribution originates from a
violent mixing in SGS}.

Carefully examining the anisotropy in the velocity distributions, 
we have also found another condition 
for the appearance of DT distributions. 
\textit{If any systematic coherent motion exists 
then it could inherit special initial condition 
and could suppress any intrinsic properties 
such as DT distribution for SGS}. 
This coherent motion actually exists 
in the radial velocity distribution 
in the spherical collapse processes, 
and in the rotational velocity distribution 
in the processes with non-vanishing angular momentum.

As a conclusion, 
we postulate that \textit{both the local virial relation and
DT velocity distribution, 
associated with the linear temperature-mass relation, 
are universal properties of SGS which undergo violent gravitational mixing}.

The origin of the above universal properties should be further investigated.
One possible candidate would be a steady heat flow from the center toward
outskirts of the spherical system. In our calculation, the two-body
relaxation process is initiated at the very center of the core region,
within our numerical calculation period. Release of the gravitational energy
at the center would yield a steady heat flow toward the outskirts of the
system. This flow may guarantee the two properties we found in this paper
and their duration. We hope we can report the detail very soon.

In this paper, we paid attention only to the velocity distribution function.
However, we need the information of matter configuration like density
profile to understand the full characteristic of gravitationally bound
systems in quasi-equilibrium state. We will show that the combination of the
local virial condition and the linear mass-temperature relation leads to a
universal density profile for gravitationally bound systems in a separate
report \cite{Sota04}. This information should give a hint for the origin of
the universality profile of dark matter or elliptical galaxies like de
Vaucouleurs law \cite{deVaucouleurs48}.

In our previous work\cite{Sota01}, we have studied the self-gravitating ring
model and have obtained a non-Gaussian velocity distribution. This velocity
distribution shows a power law and is apparently different from the results
obtained in this paper. We believe that this discrepancy originates from the
difference of the boundary conditions. In the self-gravitating ring model,
the configuration space is compact. On the other hand, in the present
simulation, the configuration space is open. We would like to extensively
consider the effect of boundary conditions soon.

In this work, 
we restricted our models only to the spherical collapses and
the cluster-pair collisions. However actual galaxies may undergo much more
diverse evolution processes including chemical evolution, the environment
effects, and a possible special nature of the cold dark matter. 
So the actual observational possibility of the universality 
we found for astronomical objects should be further investigated.

\appendix


\begin{acknowledgments}
The authers would like to thank Professor Kei-ichi Maeda for the extensive discussions.
All numerical simulations were carried out on GRAPE system at ADAC (the
Astronomical Data Analysis Center) of the National Astronomical Observatory,
Japan.
This work was supported partially by a Grant-in-Aid for Scientific
Research Fund of the Ministry of Education, Culture, Sports, Science
and Technology (Young Scientists (B) 16740152).
\end{acknowledgments}



\end{document}